\documentclass[reprint,aps,prl,superscriptaddress,showpacs]{revtex4-1}
\usepackage{bm}
\usepackage{graphicx}
\usepackage{color}
\usepackage{amsmath}
\usepackage{natbib}
\usepackage{url}
\usepackage{textcase}
\usepackage{amsthm}
\usepackage{mathcomp}
\usepackage{amsfonts}
\usepackage{amssymb}
\usepackage{mathtools}
\usepackage{physics}

\begin{document}

\title{Dynamically encircling exceptional points: Exact evolution and polarization state conversion}

\author{Absar U. Hassan}
\affiliation{CREOL/College of Optics and Photonics, University of Central Florida, Orlando, Florida 32816, USA}
\author{Bo Zhen}
\affiliation{Research Laboratory of Electronics, Massachusetts Institute of Technology, Cambridge, Massachusetts 02139, USA}
\affiliation{Physics Department and Solid State Institute, Technion, Haifa 32000, Israel.}
\author{Marin Solja{\v{c}i\'c}}
\affiliation{Research Laboratory of Electronics, Massachusetts Institute of Technology, Cambridge, Massachusetts 02139, USA}
\author{Mercedeh Khajavikhan}
\author{Demetrios N. Christodoulides}
\affiliation{CREOL/College of Optics and Photonics, University of Central Florida, Orlando, Florida 32816, USA}

\date{\today}

\begin{abstract}
We show that a two-level non-Hermitian Hamiltonian with constant off-diagonal exchange elements can be analyzed exactly when the underlying exceptional point is perfectly encircled in the complex plane. The state evolution of this system is explicitly obtained in terms of an ensuing transfer matrix, even for large encirclements, regardless of adiabatic conditions. Our results clearly explain the direction-dependent nature of this process and why in the adiabatic limit its outcome is dominated by a specific eigenstate – irrespective of initial conditions. Moreover, numerical simulations suggest that this mechanism can still persist in the presence of nonlinear effects. We further show that this robust process can be harnessed to realize an optical omni-polarizer: a configuration that generates a desired polarization output regardless of the input polarization state, while from the opposite direction it always produces the counterpart eigenstate.
\end{abstract}


\maketitle

Understanding the dynamics of time-dependent Hamiltonians is key in explaining a wide range of processes in many and diverse physical settings~\cite{TimeDepHamiltonians}. This ubiquitous class of problems is of significance since it allows one to tailor the evolution of a Hamiltonian towards certain outcomes. If a system is conservative or Hermitian, a cyclic adiabatic change in a multi-parameter space can often lead to surprising results such as for example the emergence of gauge-invariant geometric phases, as first shown by Berry~\cite{Berry-Quantal}. Of particular interest is the case where eigenvalue degeneracies are enclosed within the parameter loop. In this latter scenario, the geometric phase is robust against perturbations in the control path since it is related to the “flux” generated from the degeneracies that act as topological sources.

While the Berry phase represents an intuitive and powerful unifying notion, it is by nature based on the adiabatic theorem~\cite{Messiah}. Quite recently, a series of studies have critically reexamined these aspects in non-Hermitian environments where it was found that the system behavior can be significantly modified around degeneracies, better known as exceptional points (EPs)~\cite{Kato,*Heiss,*GarrisonWright,*EMGrafePTBoseHubbard}. As opposed to conservative systems, in non-Hermitian arrangements both the eigenvalues and the corresponding eigenvectors tend to coalesce at an EP (while unfolding associated vectors of the Jordan form)~\cite{multiparameter}. In the last few years a number of intriguing possibilities have been realized in structures supporting EPs, including loss-induced transparency~\cite{Guo}, single-mode lasing~\cite{Hodaei,*FengPTlaser}, band merging~\cite{Bo}, asymmetric diffraction~\cite{Berini} and unidirectional invisibility~\cite{ZinLin,*FengUnidirectional}, to mention a few. In other studies, the topological properties associated with the quasi-static encirclement of an EP were also investigated. Under such stationary conditions it was found that the ‘instantaneous’ eigenstates now swap with each other at the end of the parameter cycle with only one acquiring a geometric phase~\cite{HeissPhysicsofEPs,Alexei-GeometricPhase}. This behavior, attributed to the branch point character of the degeneracy that causes a gradual transition between the intersecting complex Riemann sheets, was observed in microwave cavities~\cite{Dembowski} and exciton-polariton systems~\cite{Ostrovskaya}.

This situation gets drastically altered when a non-Hermitian Hamiltonian dynamically evolves around an EP~\cite{GWunner,*Nimrod-ResonanceCoalescence,*Atabek-VibrationalCooling}. In this regime, one finds out that adiabatic predictions tend to break down~\cite{Nimrod-TimeAsymmetric,*Eva-PRAbreakdown}\textemdash a direct byproduct of the fact that the eigenvector basis is skewed and the eigenvalues themselves are generally complex. Indeed, even for slow enough cycles, numerical studies reveal that only one state dominates the output and what determines this preferred eigenstate is the sense of rotation in the parameter space~\cite{Uzdin-Ontheobservability,*Yidong-ChiralEPs,*Yidong-SciRep,*GWunnerPRA,*TLee}. These surprising effects were recently observed in microwave~\cite{Rotter-Nature} and optomechanical~\cite{Harris-Nature} systems. The unexpected transitions during such a non-Hermitian evolution have been traced to the Stokes phenomenon of asymptotics~\cite{Uzdin-Berry} and stability aspects~\cite{Rotter-PRA}. However, a full analytical treatment that systematically explains the chiral nature of the dynamics and why the system's adiabatic evolution is always funneled into a preferred eigenstate, is still lacking.
\begin{figure}
  \includegraphics{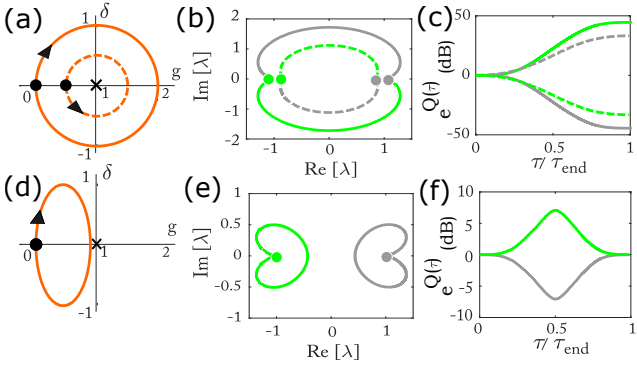}
  \caption{\label{Fig1}
  The upper [(a)-(c)] and lower [(d)-(f)] panels represent eigenvalue trajectories when the EP (marked with $\times$) is quasi-statically encircled (a) or excluded (e), respectively, from the parameter loop. Path directions are shown with arrows in (a) and (d) and black dots depict the starting points. Solid lines, throughout indicate results for a CW path and dashed, CCW. The eigenvalues ($\lambda_{1,2}$) at the start of the loop are depicted as green and gray dots in (b) and (e) where their trajectories are also shown in the corresponding colors. When the EP is enclosed, the eigenvalues swap with each other\textemdash(b) and when it is excluded, they return to themselves\textemdash(e). The accumulated gain, $e^{Q}$ (see text) corresponding to the different eigenvalue paths is plotted in (c) and (f) against time $\tau$.}
\end{figure}

In this Letter, we theoretically analyze the behavior of two coupled states whose dynamics are governed by a non-Hermitian Hamiltonian undergoing cyclic variations in the diagonal terms. The model presented here is readily realizable and even more importantly, allows one to track the modal populations at all times without imposing restrictions on the degree of adiabaticity or the size of EP encirclement. Analytical solutions obtained via confluent hypergeometric functions clearly explain the underlying asymmetric conversion into a preferred mode and the chiral nature of this mechanism is brought to the fore through appropriate transformations. Building on these findings, we propose a novel on-chip universal polarizer that can produce in a robust way a single desired output polarization irrespective of the input light state. When used in reverse, this omni-polarizer instead produces the complementary polarization eigenstate.

We here consider a system of two coupled entities, constantly exchanging energy in space or time. For example, in optics, this can be implemented using cavities, waveguides or a series of varying dichroic birefringent plates. For this $2\times2$ system, the dynamics are governed by $i\partial_t\ket{\psi(t)}+H(t)\ket{\psi(t)}=0$, where in this Schr\"{o}dinger description the time-dependent Hamiltonian $H(t)$ is given by,
\begin{equation}
 \label{full system1}
 H(t)=
\begin{pmatrix}
-i\tilde{g}(t) - \tilde{\delta}(t)   & \kappa       \\
\kappa           & i\tilde{g}(t) + \tilde{\delta}(t)
\end{pmatrix},
\end{equation}
with the state vector being $\ket{\psi(t)}=(a(t),b(t))^T$. The time (or space) varying quantities $\tilde{g}(t)$ and $\tilde{\delta}(t)$ represent gain/loss and level-detuning between the two coupled elements, respectively. Meanwhile $\kappa$ denotes the coupling strength ($\kappa\in\mathbb{R}$). The structure of Eq.~(\ref{full system1}) implies that one element [$a(t)$] is subjected to gain while the other [$b(t)$] to an equal amount of loss. For unbalanced gain-loss or detunings, one can still reduce the description to a traceless form via suitable gauge transformations~\cite{Guo,Rotter-PRA}. We henceforth use scaled variables, $(\tilde g/\kappa,\tilde{\delta}/\kappa,\kappa t)\rightarrow(g,\delta,\tau)$. In this arrangement, the EP is judiciously established in parameter space at $g=1$ and $\delta=0$, where the two coalescing eigenvalues are $\lambda_{1,2}=0$ with the corresponding eigenvectors collapsing to $\ket{\psi}=(1,i)^T$. This EP can be encircled during propagation provided that,
\begin{equation}
 \label{trajectory_cw}
g(\tau)=1-\rho\cos(\gamma\tau), \delta(\tau)=\rho\sin(\gamma\tau),
\end{equation}
where $\gamma$ is a measure of adiabaticity and $\rho$ represents the radius of the circle ($\rho \leq 1$). Equation~(\ref{trajectory_cw}) represents a clockwise (CW) loop if $\gamma>0$, and a counter-clockwise (CCW) one if $\gamma<0$. In analogy with previous studies in $\mathcal{PT}$-symmetric systems, the trajectory is chosen to start ($\tau=0$) and end ($\tau=2\pi\gamma^{-1}=\tau_{\text{end}}$) at the point that corresponds to the unbroken $\mathcal{PT}$-symmetric phase~\cite{Makris} to prevent any amplifying or decaying modes at the input-output interfaces. At these terminal points, the eigenvectors and eigenvalues are $\ket{\psi_{1,2}}=(1,\pm e^{\pm i\theta})^T$ and $\lambda_{1,2}=\pm\cos\theta$ respectively, where $\sin\theta=(1-\rho)$. The eigenvectors $\ket{\psi_{1,2}}$ are the states around which our discussion is centered. These two vectors are biorthogonal with their corresponding left eigenvectors $\ket{\tilde{\psi}_{1,2}}=(1,\pm e^{\mp i\theta})^T$.

It is instructive to first follow the motion of the system in a \textit{quasi-static} manner \footnote{Quasi-static here means that parameters are not varied continuously with time. Eigenvalues and eigenvectors are found for each set of values of (g,$\delta$) successively on the loop} by tracking the instantaneous eigenvalues. When the EP is encircled, Fig.~\ref{Fig1}(a)-(c), the eigenvalues swap with each other. And for each direction of encirclement, the path of one of them stays mostly in the negative imaginary plane. As a result, the gain-loss component of the dynamical phase $e^{Q(\tau)}$ associated with that specific eigenvalue, where $Q(\tau)=-\int_{0}^{\tau}dt'\text{Im}\left[\lambda(t')\right]$, leads to a significant amplification\textemdash Fig.~\ref{Fig1}(c). Consequently, the eigenvector that corresponds to this eigenvalue eventually dominates. On the other hand, when the loop excludes the EP, Fig.~\ref{Fig1}(d)-(f), the eigenvalues instead return to themselves and none of them is preferentially amplified at the end of the parameter excursion. This complex phase and the existence of non-adiabatic couplings between the eigenstates~\cite{Nimrod-TimeAsymmetric} eventually leads to a chiral dominance of one eigenstate over the other.

However, for a \textit{dynamical} parameter cycle a full understanding of this process can only emanate from an analytical approach. In this regard, following Eq.~(\ref{trajectory_cw}), ~(\ref{full system1}) can be re-casted into a second order differential equation for $a(\tau)$, e.g.

\begin{equation}
 \label{SecondOrder-a}
\frac{d^2a(\tau)}{d\tau^2}-\left[ \rho^2e^{2i\gamma\tau} - \rho(2+i\gamma)e^{i\gamma\tau} \right]a(\tau) = 0,
\end{equation}

A similar equation can be obtained for $b(\tau)$. We note that if the solutions corresponding to a CW loop can be obtained, they can be directly used to describe the CCW case simply by employing the transformation $(a,b)\rightarrow(a^*,-b^*)$~\cite{Suppl}. Equation ~(\ref{SecondOrder-a}) can be solved by using the substitutions $\eta=-2i\rho\gamma^{-1} e^{i\gamma\tau}$ and $a(\eta)=e^{-\eta/2}w(\eta)$, which reduce it to the form of a degenerate hypergeometric differential equation,
\begin{equation}
 \label{DegenHyp-w}
\eta\frac{d^2w(\eta)}{d\eta^2}+\left( 1-\eta\right)\frac{dw(\eta)}{d\eta} -\left(\frac{i}{\gamma}\right)w(\eta)= 0,
\end{equation}
From here one can obtain the general solution,
$a(\eta) = e^{-\eta/2}\left[c_1 F(i/\gamma,1,\eta)+c_2U(i/\gamma,1,\eta)\right]$,
where $F$ and $U$ represent confluent hypergeometric functions of the first and second kind respectively and the coefficients $c_{1,2}$ depend on initial conditions. We can now express $\begin{bmatrix} a(\tau) & b(\tau)\end{bmatrix}^T$ in the form of a transfer matrix:
\begin{equation}
 \label{ab-tau}
 \begin{bmatrix}
a(\tau) & b(\tau)
\end{bmatrix}^T
 = \sigma(\tau)M_1(\tau)M_2M_3
\begin{bmatrix}
a(0) & b(0)
\end{bmatrix}^T.
\end{equation}
Notice that only $\sigma$ and the matrix $M_1$ are $\tau$-dependent. The exact solution presented in Eq.~(\ref{ab-tau}) is general and applies regardless of adiabatic or non-adiabatic conditions and pertains to both large and small encirclements $\rho$. The scalar $\sigma$ is given by $\sigma(\tau) = i\Gamma\left(i/\gamma\right)e^{i\frac{\rho}{\gamma}\left(1+e^{i\gamma\tau}\right)}$, where $\Gamma$ is the gamma function and the matrices are described below:
\begin{subequations}\label{T-matrix}
  \begin{gather}
  M_1(\tau) =
  \begin{bmatrix}
F^{(0)}  &   U^{(0)}  \\
iF^{(0)}+\frac{2\rho}{\gamma} e^{i\gamma\tau}F^{(1)}  & iU^{(0)}-\frac{2\rho}{\gamma} e^{i\gamma\tau}U^{(1)}
\end{bmatrix}
\label{T-matrix-1},
\\
  M_2 =
  \begin{bmatrix}
-\rho U_{\tau=0}^{(0)}/\gamma-2i\rho U_{\tau=0}^{(1)}/\gamma^2  &   -U_{\tau=0}^{(0)}  \\
\rho F_{\tau=0}^{(0)}/\gamma-2i\rho F_{\tau=0}^{(1)}/\gamma^2  &   F_{\tau=0}^{(0)}
\end{bmatrix}
\label{T-matrix-2},
\\
  M_3 =
  \begin{bmatrix}
1  &   0  \\
(1-\rho)/\gamma  &   i/\gamma
\end{bmatrix}
\label{T-matrix-3}.
\end{gather}
\end{subequations}
\begin{figure}
  \includegraphics{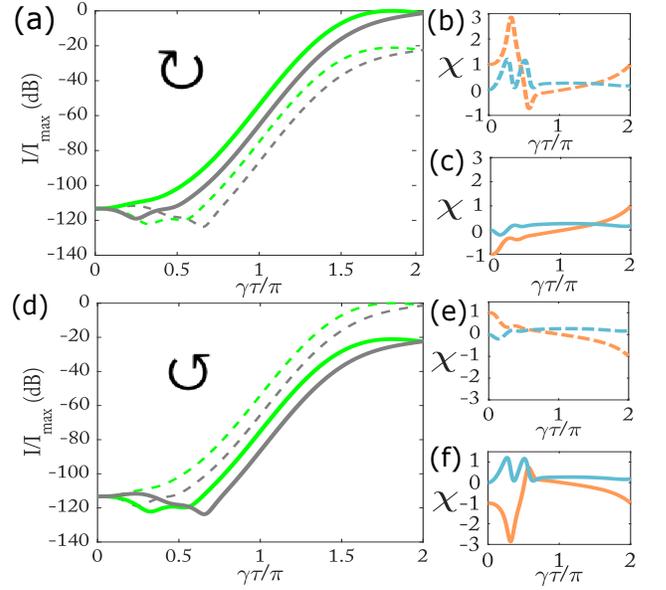}
  \caption{\label{Fig2}
  Intensity evolutions ($|a(\tau)|^2$-green, $|b(\tau)|^2$-gray) for a CW loop are shown in (a), normalized with respect to the maximum value $I_{\text{max}}$. The real (orange) and imaginary (blue) parts of the ratio $\chi(\tau)=b(\tau)/a(\tau)$ are depicted in (b) and (c). Dashed lines correspond to the input $\ket{\psi_{1}}$ and solid to $\ket{\psi_{2}}$. The same scenario for a CCW parameter loop is shown in (d)-(f). At the end of the excursion ($\tau=2\pi\gamma^{-1}$), in the CW case, Re[$\chi\rightarrow1$] and Im[$\chi\rightarrow0$] for both local eigenvector inputs, while in the CCW case Re[$\chi\rightarrow-1$] and Im[$\chi\rightarrow0$].}
\end{figure}
The abbreviated forms of the hypergeometric functions, $F^{(n)}$ and $U^{(n)}$, represent $F(n+i\gamma^{-1},n+1,-2i\rho\gamma^{-1}e^{i\gamma\tau})$ and $U(n+i\gamma^{-1},n+1,-2i\rho\gamma^{-1}e^{i\gamma\tau})$, respectively. The generic function $F(p_1,p_2,x)$ is single-valued for all complex variables $x$, where $p_{1,2}$ are complex parameters. On the other hand, the principal value of $U(p_1,p_2,x)$ is defined in the interval $-\pi<\text{arg}(x)\leq\pi$. In the case at hand, starting from $-0.5\pi$, the relevant argument, $\text{arg}(-2i\rho\gamma^{-1}e^{i\gamma\tau})$, reaches $\pi$ when $\tau=1.5\pi\gamma^{-1}$. Outside this range, i.e. for $\tau\in[1.5\pi\gamma^{-1},2\pi\gamma^{-1}]$, one has to use a connection formula according to~\cite{Slater},
\begin{multline}\label{U-analyticCont}
  U(p_1,p_2,x)=\frac{\Gamma(p_2-p_1)}{e^{i\pi p_1}}\left[\frac{F(p_1,p_2,x)}{\Gamma(p_2)}-\right. \\
  \left.\frac{e^{-i\pi(p_2-p_1)}}{\Gamma(p_1)}e^{x}U(p_2-p_1,p_2,e^{-i\pi}x)\right],
\end{multline}
By using the results of Eqs.~(\ref{ab-tau}-\ref{U-analyticCont}), the intensity evolution ($|a|^2,|b|^2$) in the two coupled entities is shown in Fig.~\ref{Fig2} when $\gamma=0.4$ and $\rho=1$, i.e. $\theta=0$. Figures 2(a)-(c) depict the CW case whereas 2(d)-(f) show similar results for a CCW scenario. By monitoring both the real and imaginary components of the modal fields, one finds out that both local eigenvectors $\ket{\psi_{1,2}}$ at $\tau=0$ are eventually transformed at the end of the cycle to $\ket{\psi(\tau_{\text{end}})}\propto\ket{\psi_1}=(1,e^{i\theta})^T$ if the loop is performed in a CW fashion\textemdash Figs. 2(b) and 2(c). Of course this is also true for any linear combination of the two eigenvectors at the input. Conversely, if the encirclement is carried out in a CCW manner, $\ket{\psi(\tau_{\text{end}})}\propto\ket{\psi_2}=(1,-e^{-i\theta})^T$ for any input state\textemdash Figs. 2(e) and 2(f). To understand this chiral mode preference mechanism, we take a closer look at the elements $\boldsymbol{m_{ij}}$ of the transfer matrix $M(\tau)=M_1(\tau)M_2M_3$ by considering the complex ratio $\chi(\tau)=b(\tau)/a(\tau)$.
\begin{equation}\label{b_by_a-Hyp}
  \chi(\tau)=\frac{\boldsymbol{m_{21}}a(0)+\boldsymbol{m_{22}}b(0)}{\boldsymbol{m_{11}}a(0)+\boldsymbol{m_{12}}b(0)}
\end{equation}

The aforementioned mode-conversion is evidently only possible under adiabatic conditions, $\gamma\ll1$. In this regime, one finds a very specific proportionality factor between pairs of $\boldsymbol{m_{ij}}$. Based on analytic continuation at $\tau=\tau_{\text{end}}$, the asymptotic behavior of the matrix elements leads to the following important relation, $\boldsymbol{m_{21}}/\boldsymbol{m_{11}}=\boldsymbol{m_{22}}/\boldsymbol{m_{12}}=i+(2\rho F_{\tau=0}^{(1)})/(\gamma F_{\tau=0}^{(0)})$~\cite{Suppl}. Given that the terms $i\gamma^{-1}$ and $-2i\rho\gamma^{-1}$ are both large for $\gamma\ll1$, we now use the asymptotic expansion of $F(p_1,p_2,x)$ for large $p_1$~\cite{Slater},
\begin{multline}\label{F_aprx}
  F(p_1,p_2,x)\sim \Gamma(p_2)e^{x/2}(kx)^{(1-2p_2)/4} \\
  \pi^{-1/2}\cos(2\sqrt{kx}-\pi p_2/2+\pi/4),
\end{multline}
where $k=p_2/2-p_1$. As a result, $F_{\tau=0}^{(1)}/F_{\tau=0}^{(0)}\sim\gamma(e^{i\theta}-i)/(2\rho)$. From here, we finally obtain:
\begin{equation}\label{Conversion-Hyp}
  \frac{\boldsymbol{m_{21}}}{\boldsymbol{m_{11}}}=\frac{\boldsymbol{m_{22}}}{\boldsymbol{m_{12}}}\sim e^{i\theta}.
\end{equation}

Therefore, $\chi(\tau_{\text{end}})=b(\tau_{\text{end}})/a(\tau_{\text{end}})\rightarrow e^{i\theta}$. Equations~(\ref{b_by_a-Hyp},\ref{Conversion-Hyp}) lie at the heart of our results since they explain in a comprehensive manner why this mode-conversion process takes place, with all possible inputs converging towards $\ket{\psi_{1}}=(1,e^{i\theta})^T$. Moreover, given the fact that $(a,b)\rightarrow(a^*,-b^*)$ for a counterclockwise loop, while leaving the matrix elements unchanged, one directly finds that for a CCW encirclement, $\chi(\tau_{\text{end}})\rightarrow -e^{-i\theta}$ and as a result all possible inputs are converted into $\ket{\psi_{2}}=(1,-e^{-i\theta})^T$. An interesting subcase of our results arises when $\rho\ll1$, where the hypergeometric functions can be accurately represented by Bessel functions~\cite{Suppl}.
\begin{figure}
  \includegraphics{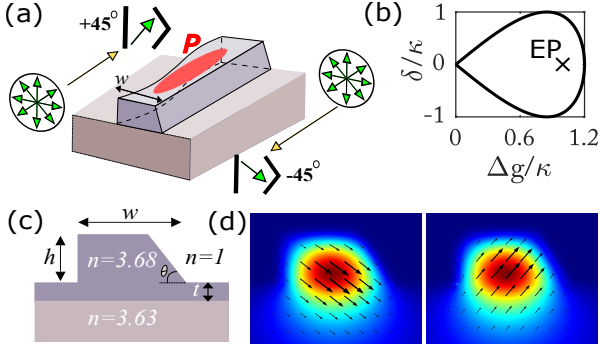}
  \caption{\label{Fig3}
  (a) A possible realization of an omni-polarizer is shown that highlights the variations in the width\textemdash$w$ (sinusoidal) and pumping\textemdash$P$ (strongest in the center). Direction-dependent polarization conversions are also schematically illustrated with green arrow-heads. (b) To limit the required maximum amplification, the parameter loop around the EP ($\times$) is here chosen to be skewed. Detuning is given by $\delta$ and $\Delta g$ represents the difference between TE and TM modal gains, i.e. $\Delta g=(g_x-g_y)/2$. (c) A cross-section at $z=0,L/2$ and $L$ ($L$\textemdash length of the device) is shown where the dimensions are $(h,w,t)=(0.8,1.42,0.1)$ $\mu$m and $\theta=70^{\circ}$. In this system $h, t$ and $\theta$ are kept constant throughout, while $w$ varies as, e.g. $w=1.42-0.08\sin(2\pi z/L)$. The refractive indices for this GaAs-AlGaAs structure are also shown in (c) at the wavelength of 800 nm. (d) The two resulting orthogonal eigenmodes with their electric field polarizations.}
\end{figure}

In many applications it is often required to control the polarization state at the output of a system~\cite{MWatts,Reano-Berry}. In particular, significant effort has been invested in overcoming the polarization dependent performance of components such as optical amplifiers and wavelength filters. Based on the results presented earlier, we here propose a single channel omni-polarizer. This structure is expected to transform any input into a desired state of polarization ($\ket{\psi_{1}}$) when light traverses it in one direction. Conversely, in the opposite direction, any arbitrary input is mapped into the biorthogonal vector ($\ket{\psi_{2}}$). A possible realization is shown in Fig. 3(a). In this case, the slanted side-wall allows for coupling ($\kappa$) between the TE ($\hat{x}$) and TM ($\hat{y}$) polarizations~\cite{BARahman} while variations in the width of the waveguide $w$ can introduce a variable birefringence. To achieve the aforementioned conversion of any input to a single polarization, $w$ and the amount of carrier injection $P$ (optical or electrical) need to be varied along propagation so as to encircle the EP, as shown for example in Fig. 3(b). Note that $w$ and $P$ are directly related to $\delta$ and $\Delta g$ respectively. The TE and TM polarization gains vary linearly with $P$, only the latter being less by a factor of $\varepsilon$, typically $\varepsilon\sim1/3$. A cross-sectional view of this structure is shown in Fig.~\ref{Fig3}(c) where the birefringence ($\delta$) is negligible and the eigenpolarizations ($\pm 45^{\circ}$) are $\ket{\psi_{1,2}}=(1,\pm 1)^T$\textemdash Fig.~\ref{Fig3}(d).
\begin{figure}
  \includegraphics{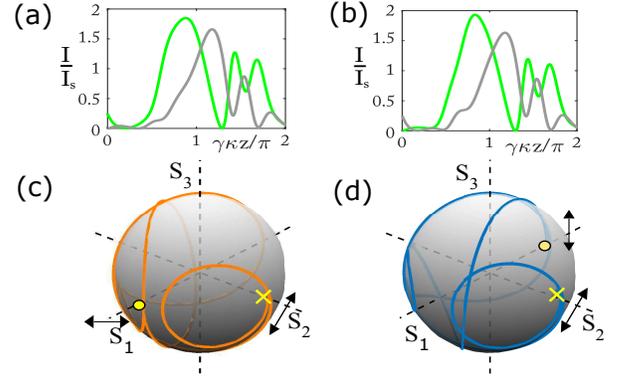}
  \caption{\label{Fig4}
  Evolution of intensities ($|E_x|^2$\textemdash green and $|E_y|^2$\textemdash gray) for the nonlinear system are shown in (a) and (b) corresponding to a TE and TM input respectively. The results are scaled with respect to the saturation intensity ($I_s$) of the gain medium. Polarization dynamics on the Poincar\'e sphere corresponding to these two cases are depicted in (c) and (d) where yellow dots indicate the input light state and crosses that of the output.}
\end{figure}

In the design presented, $\kappa$ remains nearly constant at $\kappa\sim1.4\times10^{3}\text{ m}^{-1}$. The corresponding value of maximum gain (intensity-wise) required is $100\text{ cm}^{-1}$ near the middle of the device. Meanwhile $w$ needs to be gradually varied according to $w=1.42\rightarrow1.34\rightarrow1.50\rightarrow1.42\text{ }\mu\text{m}$, as schematically shown in Fig.~\ref{Fig3}(a). Here the nonlinear evolution dynamics are given by~\cite{Suppl,Agrawal-TETM}:
\begin{subequations}\label{saturable_system}
\begin{alignat}{2}
 \frac{dE_x}{dz} = \frac{g_{x}E_x}{1+|E_x|^2+\varepsilon|E_y|^2} - (\alpha+i\delta)E_x + i\kappa E_y, &\\
 \frac{dE_y}{dz} = \frac{\varepsilon g_{x}E_y}{1+|E_x|^2+\varepsilon|E_y|^2}- (\alpha-i\delta)E_y + i\kappa E_x, &
\end{alignat}
\end{subequations}
where $\alpha\sim0.9\kappa$ is a linear absorption loss, $\gamma$ is here chosen to be $\gamma=0.4$ corresponding to a device length of $L=1.1\text{ cm}$. For a CW loop shown in Fig. 3(b), gain and detunings are dynamically varied as $g_x=3.6\kappa\sin(\gamma\kappa z/2)$ and $\delta=\kappa\sin(\gamma\kappa z)$ for $\kappa z\in[0,2\pi\gamma^{-1}]$. The ensuing evolution of intensities $|E_x|^2$ and $|E_y|^2$, scaled with the saturation level~\cite{Suppl}, is shown in Fig. 4(a) for a TE and in Fig. 4(b) for a TM input. Unlike the linear case studied before, here the intensities evolve within reasonable limits due to saturation effects. The nature of the underlying polarization conversion is revealed in Figs.~\ref{Fig4}(c) and ~\ref{Fig4}(d) where the corresponding field trajectories are plotted on the Poincar\'e sphere. Clearly, both TE and TM polarizations end up in the same eigenstate, i.e. $+45^{\circ}$ linearly polarized. For a CCW traversal, viz. $\delta=-\kappa\sin(\gamma\kappa z)$, the output polarization was found to be $-45^{\circ}$. Our results indicate that despite the presence of nonlinearities, the chiral mechanism of mode-preference still persists. In other words, the topological nature of EP encircling in this omni-polarizer makes it highly robust.

In conclusion we have provided an analytic explanation of the chiral mode-conversion mechanism that takes place during \textit{dynamic} encirclement of an EP. We demonstrated that this effect can be exploited to implement an optical omni-polarizer that exhibits counter-intuitive polarization properties. Finally, this EP based mechanism could find manifestations in various settings beyond optics, e.g. coherent population control between coupled energy levels and other non-Hermitian acoustic~\cite{CTChan-acoustics} and electronic~\cite{KottosElect} systems.

\begin{acknowledgements}
A.U.H. and D.N.C. were partly supported by NSF (Grant no. DMR-1420620) and AFOSR MURI (Grant no. FA9550-14-1-0037). M.K. acknowledges financial support from ARO (Grant nos. W911NF-16-1- 0013, W911NF-14-1-0543), ONR (Grant no. N00014-16-1-2640) and NSF (Grant no. ECCS-1454531). M.S. was partly supported by the Army Research Office through the Institute for Soldier Nanotechnologies under contract no. W911NF-13-D-0001. B.Z. was partially supported by the United States-Israel Binational Science Foundation (BSF) under award no. 2013508.
\end{acknowledgements}

\bibliography{References}

\end{document}